\documentclass{moriond}

\usepackage{amsmath}
\usepackage{bm}

\def\Journal#1#2#3#4{{#1} {\bf #2}, #3 (#4)}

\newcommand{\Photo}{\includegraphics[height=35mm]{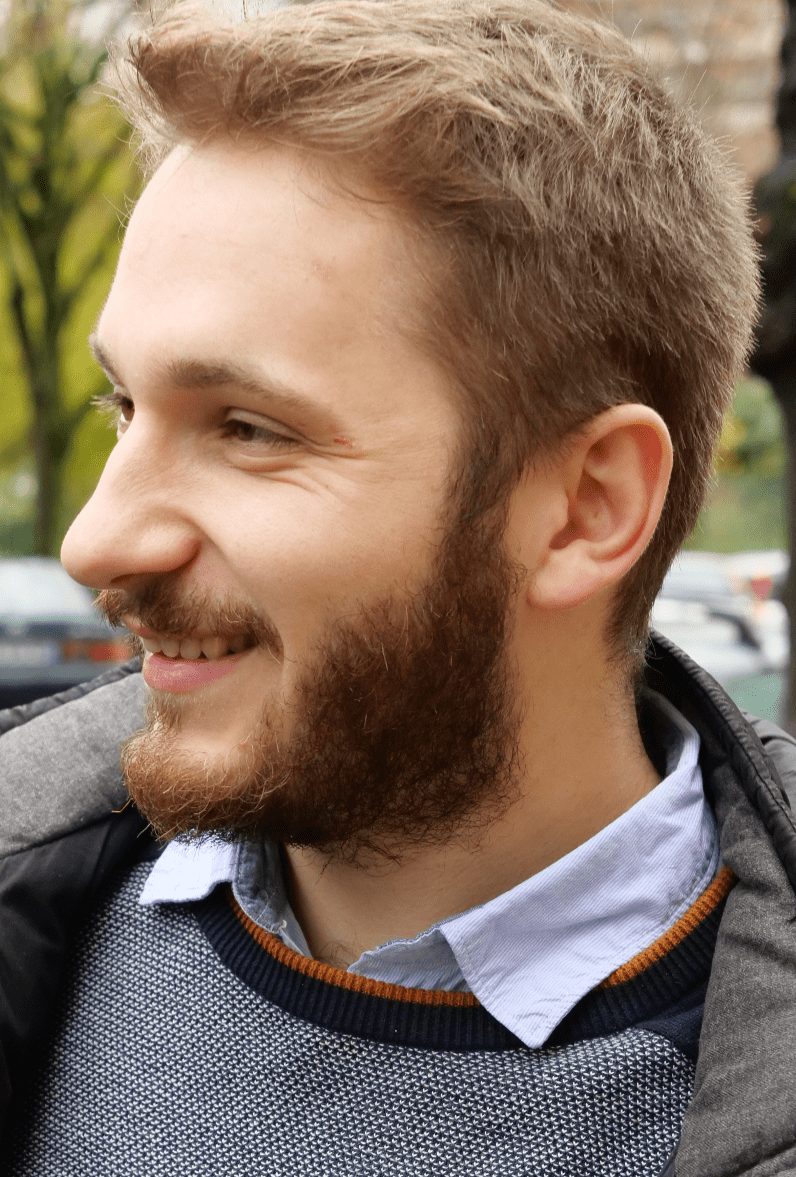}}

\begin{document}
\vspace*{4cm}
\title{One-dimensional power spectrum from first DESI Lyman-$\bm{\alpha}$ forest}

\author{ Corentin Ravoux \\
On behalf of the DESI Lyman-alpha working group}

\address{Université Clermont-Auvergne, CNRS, LPCA, 63000 Clermont-Ferrand, France}

\maketitle\abstracts{
The Lyman-alpha forest is a unique probe of large-scale matter density fluctuations at high redshift $z>2$. We measure the one-dimensional Lyman-alpha forest power spectrum using the first data provided by the Dark Energy Spectroscopic Instrument (DESI), with a fast Fourier transform estimator\cite{desi}. The data sample contains quasar spectra included in the DESI Early Data Release and the first two months of the main survey. This first set of data already provides an improvement in terms of spectroscopic resolution with respect to the previous measurements. We investigated methodological and instrumental contaminants associated with DESI and used synthetic data to validate and correct our measurement. Coupling our measurement with theoretical predictions from hydrodynamical simulations will yield strong constraints on the primordial matter power spectrum, neutrino masses, and dark matter properties. A quadratic maximum likelihood estimator was applied to the same data set on a companion paper\cite{qmle} and agrees with our measurement.}

\section{One-dimensional power spectrum}

The Lyman-alpha (Ly$\alpha$) forest is a tracer of neutral hydrogen in the cosmic web, observed in quasar spectra. When observed from ground-based telescopes at redshifts $z>2$, quasar spectra show a broad peak of Ly$\alpha$ emission at $\lambda_{\rm rest} = 1215$\AA. Bluewards of this peak, a forest of lines corresponds to light absorption by the intergalactic medium (IGM) located between the quasar and the observer. These absorption features constitute the Ly$\alpha$~forest, and trace the neutral hydrogen in the IGM. As a first step, the product of the continuous emission of the quasar $C_{q}$ by the average fraction of transmitted flux $\overline{F}$ is measured. From a quasar flux $f(\lambda)$, it is then possible to define the Ly$\alpha$~absorption contrast\cite{desi} as

\begin{equation}
    \label{eq:delta_flux_continuum}
    \delta_{F}(\lambda) = \frac{f(\lambda)}{C_{q}(\lambda_{\mathrm{rf}})\overline{F}(\lambda)} - 1\,.
\end{equation}

This contrast is the starting point for all correlation calculations with Ly$\alpha$~forests. The most standard analysis is the measurement of baryon acoustic oscillation signal on the Ly$\alpha$~auto-correlation at large scales (see e.g.\cite{bao}). In parallel, the small-scale distribution of neutral hydrogen ($\sim$ Mpc), imprinted in the fluctuations of the Ly$\alpha$~forest along the line-of-sight, is accessible by measuring the one-dimensional Ly$\alpha$~forest power spectrum in Fourier space (denoted $P_{1\mathrm{D},\alpha}$). We compute the one-dimensional Ly$\alpha$~power spectrum with the Fast Fourier Transform (FFT) method, for which $P_{1\mathrm{D},\alpha}$ is defined as the ensemble average of the squared Ly$\alpha$~contrasts in Fourier space, i.e. $P_{1\mathrm{D},\alpha}(k) = \left\langle|\delta_F(k)|^2\right\rangle$.

The one-dimensional power spectrum is sensitive to the amplitude and slope of the matter power spectrum at redshift $z>2$ and for small scales ($k \sim 1\, \mathrm{Mpc}^{-1}$). $P_{1\mathrm{D},\alpha}$ measurements are also sensitive to the thermal state and history of the IGM, which can appear as contaminants when focusing on cosmological constraints. Due to its sensitivity to the matter fluctuations at small scales, $P_{1\mathrm{D},\alpha}$ measurements can constrain physics beyond the Standard Model, such as the mass of neutrinos, the mass of warm dark matter candidates, or a possible running of the spectral index due to primordial inflation physics. The previous intermediate-resolution measurement of $P_{1\mathrm{D},\alpha}$ was performed on eBOSS data\cite{eboss}. The latest constraints\cite{palanque} were obtained by combining this measurement with Planck cosmic microwave background (CMB) for neutrino mass, and with high-resolution $P_{1\mathrm{D},\alpha}$ measurements for warm dark matter mass.

\section{DESI measurement}

The Dark Energy Spectroscopic Instrument (DESI)\cite{desioverview} is a multi-object spectrograph with $5,000$ robotically controlled fibers, which starts its main survey in June 2021. In our study, we use a specific data set of the DESI early data release called the 'one-percent survey' and noted SV3. We also use the first two months of data, named DESI-M2. The final data set used is obtained by coupling those two samples and is noted SV3+M2. After applying quality cuts on the DESI quasar catalog, we get 26,330 quasar spectra whose Ly$\alpha$ forest range from redshift $z=2.1$ to $z=3.8$. We improve the reshift resolution of our measurement by cutting each forest into three equal-sized sub-forests. The total data set contains 73,839 sub-forests.

\begin{figure}
\begin{minipage}{1.0\linewidth}
\centerline{\includegraphics[width=1.0\textwidth]{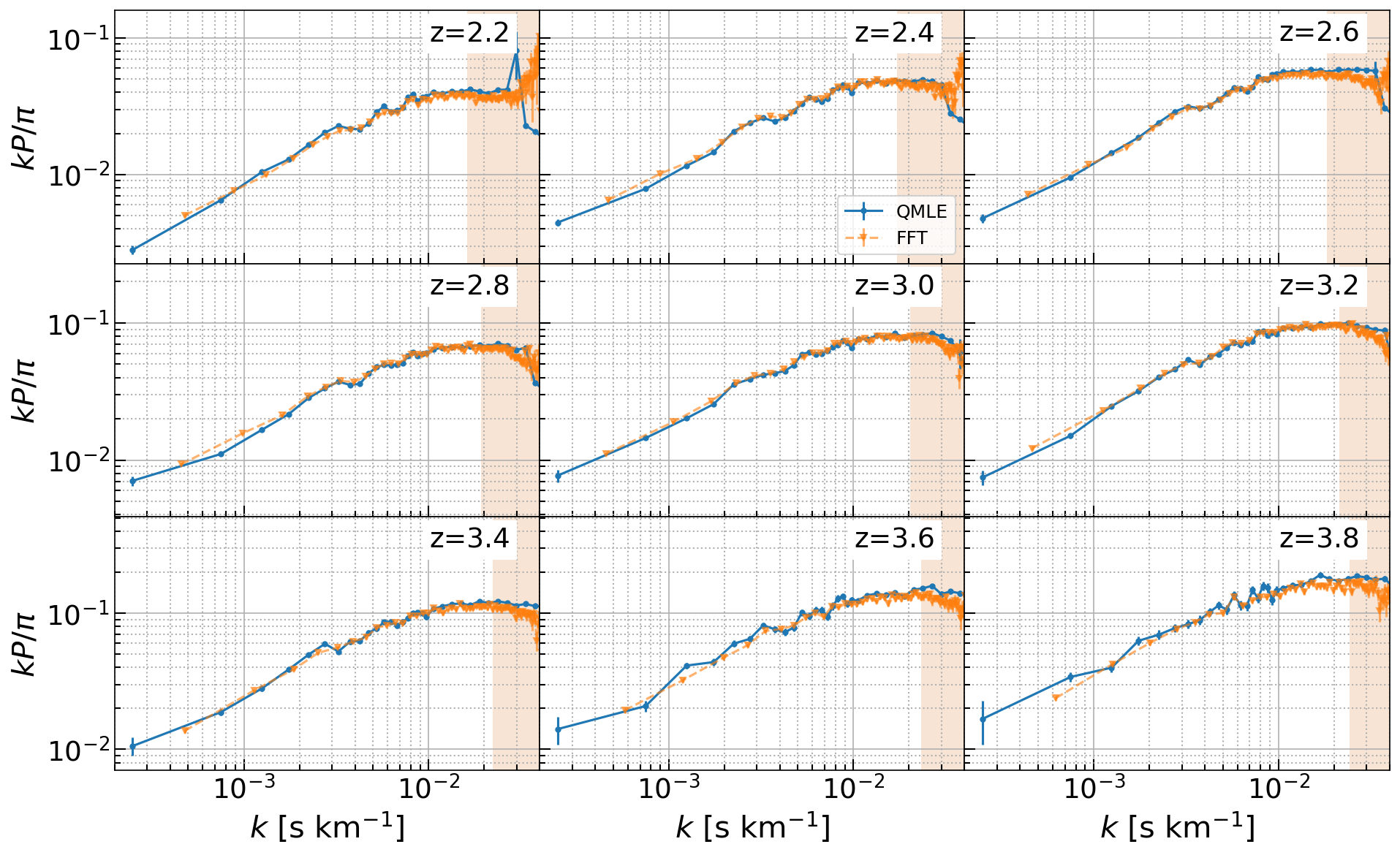}}
\end{minipage}
\caption{Normalized one-dimensional Ly$\alpha$~forest power spectrum ($kP_{1\mathrm{D},\alpha}(k)/\pi$) using the SV3+M2 data set, for redshift bins from $z=2.2$ to $z=3.8$. The points in orange corresponds to our FFT measurement, and the blue points to the QMLE measurement presented in the companion paper.}
\label{fig:p1d}
\end{figure}

In practice, the one-dimensional power spectrum is impacted by many different Ly$\alpha$ contaminants present in different places of the Universe. As mentioned in Eq.~\ref{eq:delta_flux_continuum}, we have to account for the quasar continuous emission, also called the continuum. Some quasars show specific absorption lines associated with absorptions in the accretion disk of the quasar; they are called Broad Absorption Line (BAL) quasars. By traveling between the quasar and the observer, the light can be absorbed by other IGM elements than neutral hydrogen, called metals. In its path, the light can cross higher-density regions of the circumgalactic medium. It leads to extended saturated absorption called Damped Ly$\alpha$ (DLA) systems, which contaminate the spectrum. Finally, several effects appear near the telescope. We need to account for atmospheric emission lines, the instrument's noise, and the spectrograph's finite resolution. The final FFT $P_{1\mathrm{D},\alpha}$ estimator is given by

\begin{equation}
\label{eq:final_p1d_estimator}
\begin{aligned}
&P_{1\mathrm{D},\alpha}(k) = A_{\mathrm{line}}(z,k) \cdot A_{\mathrm{dla}}(z,k) \cdot A_{\mathrm{cont}}(z,k) \cdot A_{\mathrm{res}}(k) \cdot \\
&~\left(\left\langle\left[ |\delta_F(k)|^2 -P_{\mathrm {noise}}(k) \right] \cdot \mathbf{R}^{-2}(k) \right\rangle - P_{\mathrm{metals}}(k) \right)\,.
\end{aligned}
\end{equation}

The derivation of this estimator's terms is presented in detail in the main article\cite{desi}. We performed several studies in order to fit the quasar continuum, to detect and remove BAL quasar, to detect and mask DLA objects, to compute the metal power spectrum $P_{\mathrm{metals}}$, to mask atmospheric emission lines, to measure the noise power spectrum $P_{\mathrm{noise}}$, and to correct the finite spectroscopic resolution with the $\mathbf{R}$ matrix. Using dedicated synthetic data, we also correct the masking of the atmospheric lines ($A_{\mathrm{line}}$), the masking of DLA objects ($A_{\mathrm{dla}}$), the impact of continuum fitting ($A_{\mathrm{cont}}$) and the miss modeling of resolution ($A_{\mathrm{res}}$). 

\begin{figure}
\begin{minipage}{0.49\linewidth}
\centerline{\includegraphics[width=1.0\linewidth]{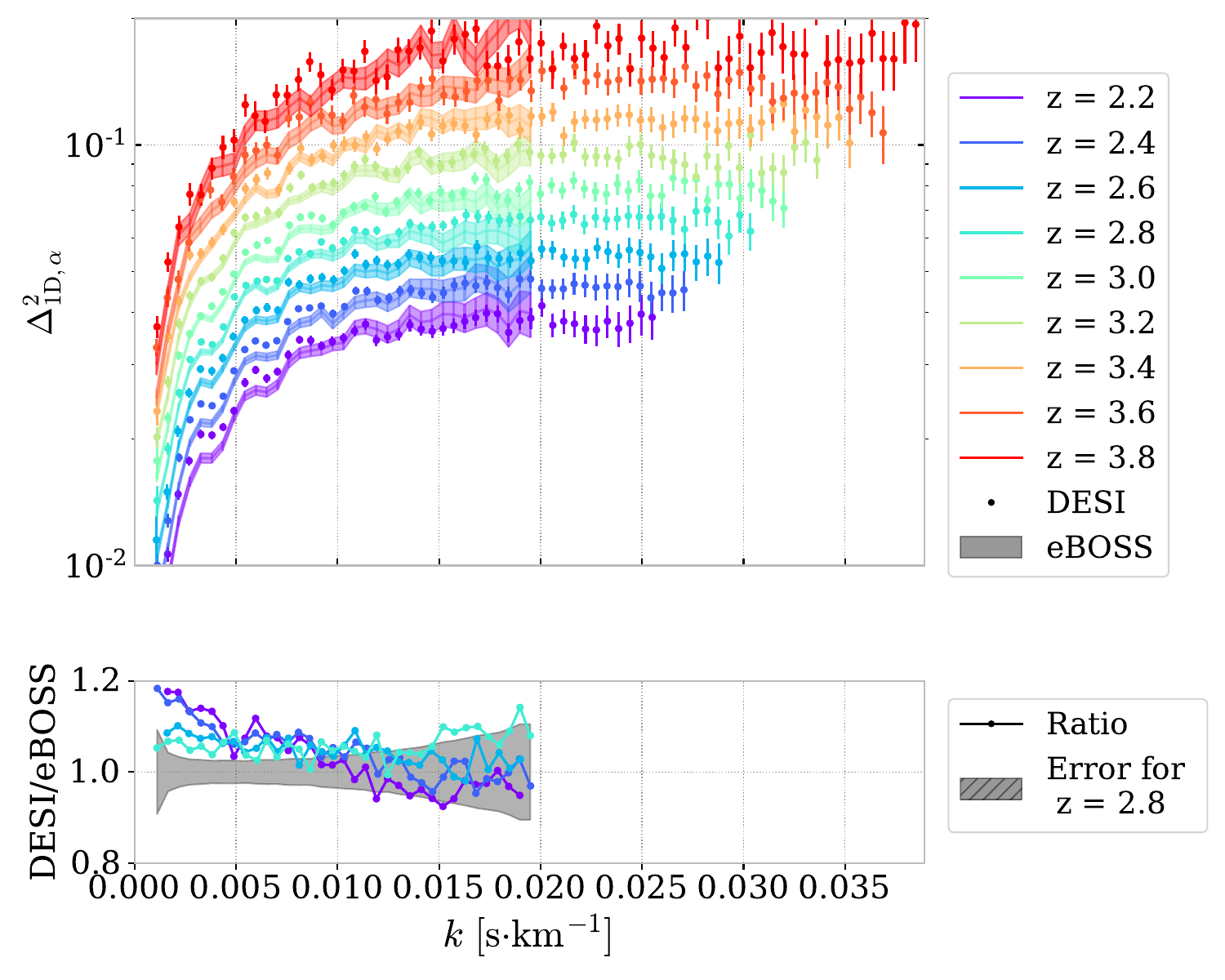}}
\end{minipage}
\hfill
\begin{minipage}{0.49\linewidth}
\centerline{\includegraphics[width=1.0\linewidth]{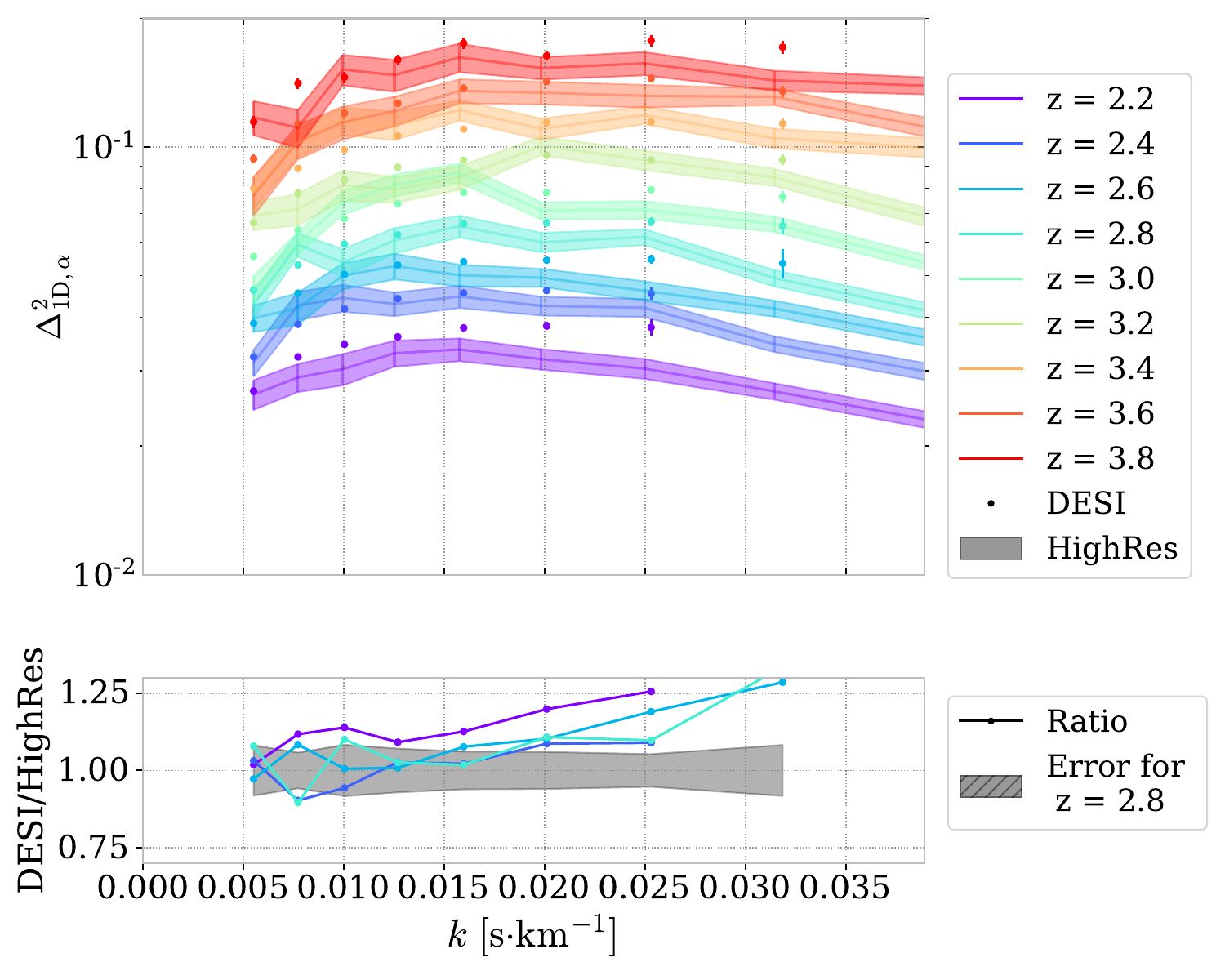}}
\end{minipage}
\caption{(left) Comparison with the eBOSS measurement. The ratio between our measurement and the eBOSS is shown in the bottom panel. The striped gray area in the bottom panel shows the centered error bar of the ratio averaged over all shown redshift bins. (right) Same comparison with the high-resolution measurement. In this case, our $P_{1\mathrm{D},\alpha}$ measurement and the error bars associated are rebinned to the wavenumber binning of the high-resolution measurement.}
\label{fig:p1d_comp}
\end{figure}

The result of applying this FFT estimator to the DESI SV3+M2 data sets is shown in Fig.~\ref{fig:p1d} (taken from\cite{qmle}). To complement this measurement, a quadratic maximum likelihood estimator (QMLE) was applied to the same dataset and presented in a companion paper\cite{qmle}. Both measurements, represented in the figure for different redshift bins, agree for all scales. The uncertainties associated with both measurements combine the statistical error bars derived from the standard deviation between the different Ly$\alpha$ forests and the systematical uncertainties we computed for each previously mentioned contaminant. The Fig.~\ref{fig:p1d_comp} shows the comparison of our FFT $P_{1\mathrm{D},\alpha}$ measurement with the previous eBOSS measurement\cite{eboss} and the latest high-resolution measurement on the combination of KODIAQ, SQUAD, and XQ-100 surveys\cite{hr}. Our measurement shows a good agreement with both measurements on intermediate wavenumbers. However, we have a disagreement with the eBOSS measurement on the largest scales. This disagreement is being investigated, and we think it is associated with incompleteness in the used DLA catalog.

\section{Cosmological interpretation and forecasts}

\begin{figure}
\begin{minipage}{0.49\linewidth}
\centerline{\includegraphics[trim=0cm 0cm 5cm 5cm, clip=true, width=1.0\linewidth]{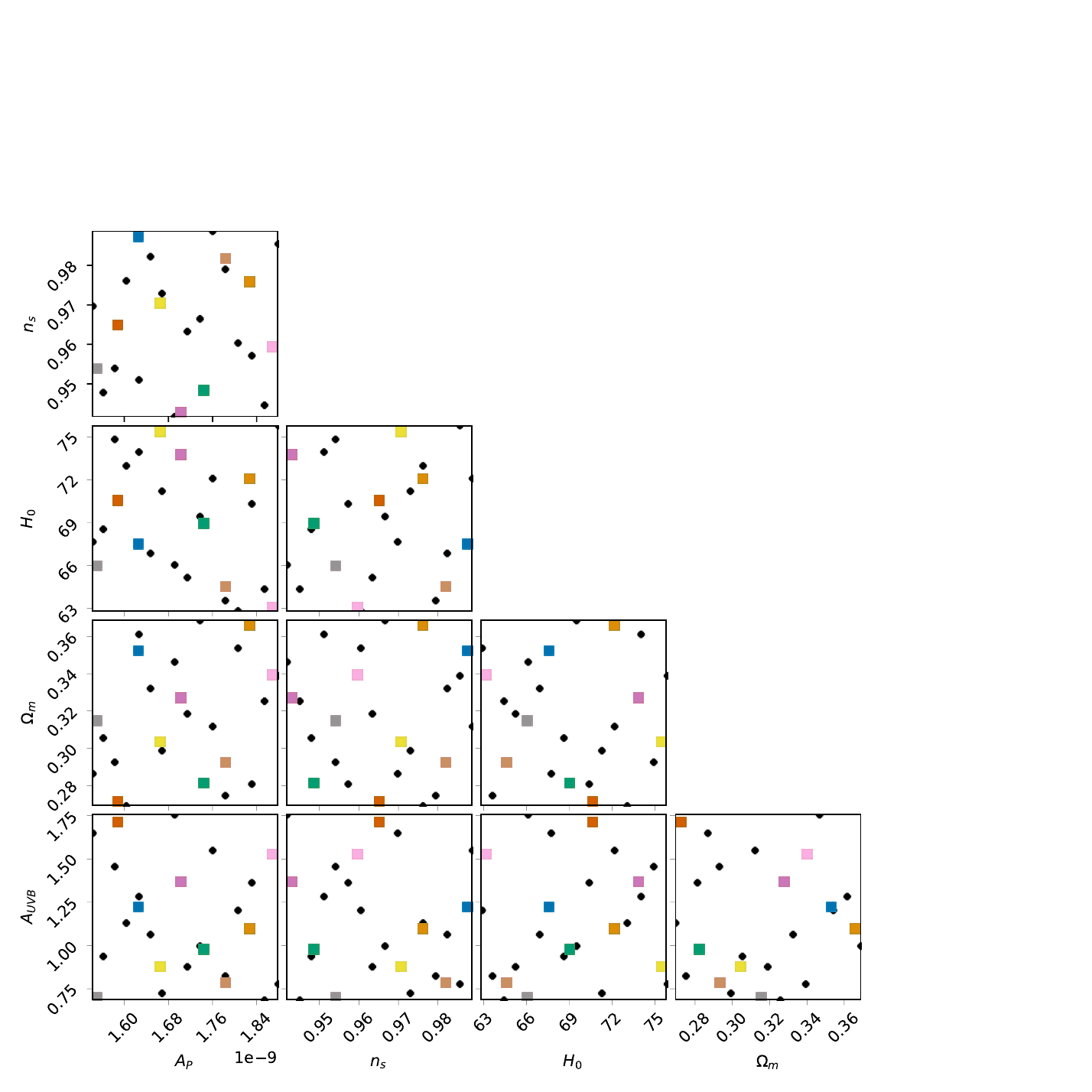}}
\end{minipage}
\hfill
\begin{minipage}{0.49\linewidth}
\centerline{\includegraphics[width=1.0\linewidth]{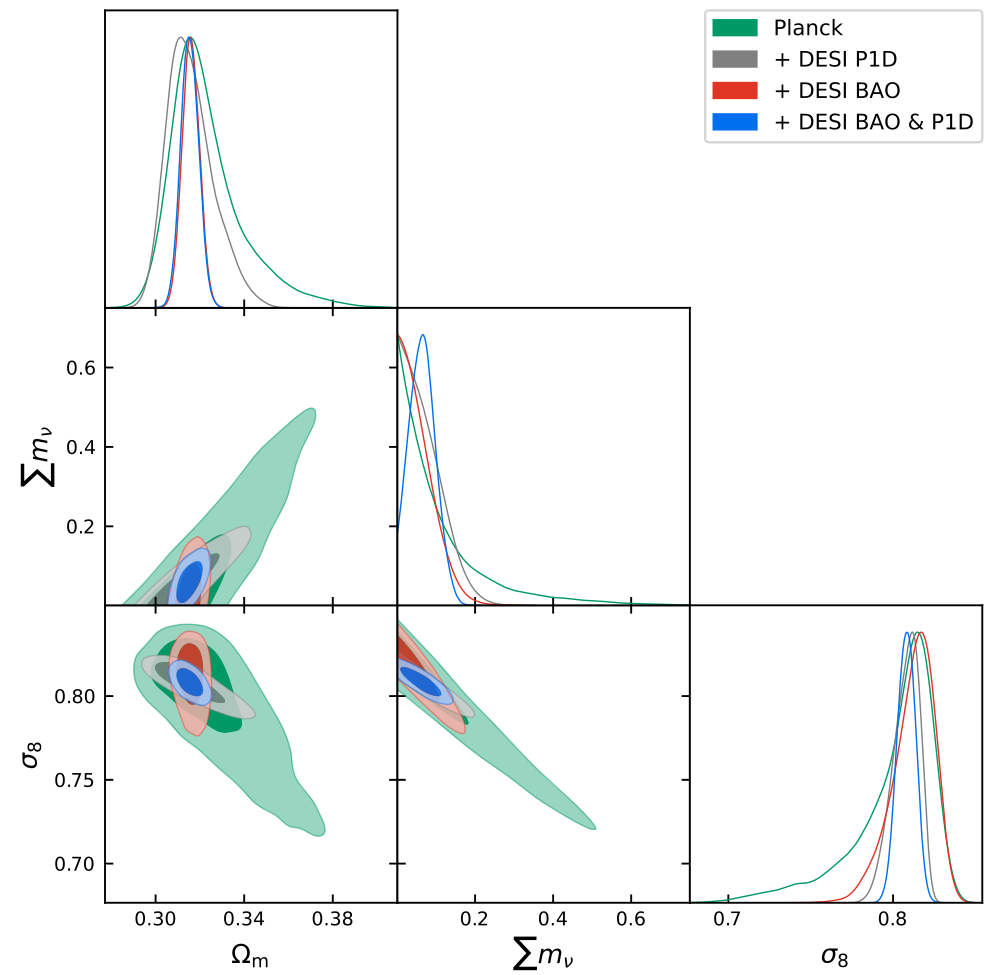}}
\end{minipage}
\caption{(left) An example of cosmological parameters chosen to cover the parameter space uniformly. A hydrodynamic simulation and a $P_{1\mathrm{D},\alpha}$ prediction are performed for each parameter. A Gaussian process emulator is trained on this set of simulations. (right) Fisher forecast of the constraints achievable by the final DESI $P_{1\mathrm{D},\alpha}$ measurement in association with Planck CMB (grey) and Planck CMB+DESI BAO (blue). In comparison, the constraints obtained by Planck alone are shown in green and Planck CMB+DESI BAO in red.}
\label{fig:interp}
\end{figure}

Because $P_{1\mathrm{D},\alpha}$ probes very small scales in the IGM, its cosmological interpretation necessitates using very high-resolution hydrodynamic simulations. Those simulations are very costly in computation time, and it is impossible to launch a simulation for each needed theoretical estimation of $P_{1\mathrm{D},\alpha}$. We choose to develop emulators that estimate the $P_{1\mathrm{D},\alpha}$ for each cosmological and IGM thermal parameter. This kind of emulator is trained on a small number of simulations uniformly distributed in the cosmological parameter space, such as the left panel of Fig.~\ref{fig:interp} for the emulator taken from\cite{hydro}. Those emulators can predict the theoretical value of $P_{1\mathrm{D},\alpha}$ at the 1\% level for scales concerned by DESI. Considering the disagreement between our measurement and eBOSS and the complexity of interpreting $P_{1\mathrm{D},\alpha}$ measurement, we decided not to perform the cosmological interpretation. However, the right panel of Fig.~\ref{fig:interp} shows a Fisher forecast on the cosmological constraints that the $P_{1\mathrm{D},\alpha}$ will be able to achieve at the end of the DESI survey. In association with BAO measurement from DESI and CMB probes, we aim to achieve a measurement of the sum of neutrino mass with an accuracy of $\sigma \left( \sum m_{\nu} \right) = 0.03 \mathrm{eV}$.

\end{document}